\documentclass[twocolumn,showpacs,preprintnumbers,amsmath,amssymb]{revtex4}

\usepackage{graphicx}
\usepackage{dcolumn}
\usepackage{bm}
\usepackage[colorlinks]{hyperref}

\begin{document}


\title{Mechanisms of recoverable prevalence and extinction of viruses\\ on linearly growing scale-free networks}

\author{Yukio Hayashi}
\affiliation{%
Japan Advanced Institute of Science and Technology,\\
Ishikawa, 923-1292, Japan
}%


\date{\today}

\begin{abstract}
We investigate mechanisms of the typically observed recoverable 
prevalence in epidemic spreading.
Assuming the time-independent connectivity correlations, 
we analyze the dynamics of 
spreading on linearly growing scale-free (SF) networks, and derive the 
extinction condition related to the rates of network growth, infection, 
and immunization of viruses.
The behavior is consistent with the previous results for SF networks 
by a mean-field approximation without connectivity correlations.
In particular, it is suggested that 
the growing must be stopped to prevent the spreading of 
infection.
This insight 
helps to understand the spreading phenomena on 
communication or social networks.
\end{abstract}

\maketitle


In the world-wide communication and transportation, we are able to work
on rapid consideration or decision for scientific, technological,
social, and business trends, while we are also frighten by threat to
dangerous viruses for computer systems and our body itself.
The needs become larger for 
understanding the mechanism of epidemic spreading and for 
finding effective strategies of the prevention against huge damage in 
economy or life by computer viruses via e-mails
\cite{Kephart93b} \cite{Trendmicro03}, 
human immunodeficiency virus (HIV) \cite{Liljeros01}, 
severe acute respiratory syndrome (SARS) \cite{Lipsitch03},
and so on. 
For the purposes, the study of infection disease on network models is 
useful. In particular, we should remark that    
many real networks, whose node and link are  
corresponding to individual (computer or person) and the  
(communicational, pathological, or sexual) contact between them,  
have scale-free (SF) structures based on natural rules of network 
growth with 
new nodes and preferential attachment for linking  
as rich-get-richer phenomenon \cite{Albert01} \cite{Barabasi99},  
in which the degree distribution  
exhibits a power-law as $P(k) \sim k^{-\gamma}$, $2 < \gamma < 3$,   
for the probability of $k$ connections in address-books of  
e-mails \cite{Ebel02} \cite{Hayashi03}  
or sexual partners \cite{Liljeros01}.

It is very possible that there exist a common mechanism for epidemic  
spreading in realistic SF networks,   
and that the extremely heterogeneity consisted of many  
normal nodes with less connectivities and a few hubs with much
connectivities is crucial for the superspreading \cite{Lipsitch03}
and the persistence of viruses in long-time \cite{Trendmicro03}.
Indeed, the important properties of SF networks are the robust and 
vulnerable connectivities against
random and targeted attacks, respectively \cite{Albert00a}.
The property is applicable for the targeted immunization for hubs
\cite{Dezso02} \cite{Satorras02}
to prevent the spreading of infection.
Thus, the universal and robust structure in growing complex networks
attracts physicists, 
and surprising results have been recently obtained.

\begin{figure}[htb]
    \includegraphics[width=70mm]{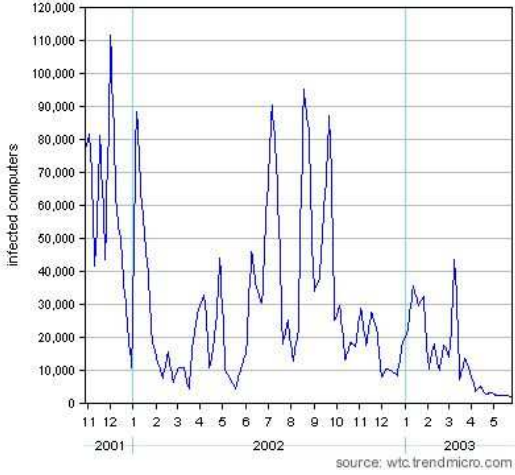} 
    \caption{Typically observed pattern as recoverable prevalence of the
 computer virus NIMDA extracted from \cite{Trendmicro03}.}
    \label{fig_NIMDA}
\end{figure}

In striking contrast with the usual models for epidemic
spreading \cite{Anderson92},
it has been shown \cite{Satorras01a} that
a susceptible-infected-susceptible (SIS) model on heterogeneous 
SF networks has no epidemic threshold at a large network size nearly
infinity; 
infection can be proliferated, whatever
small infection rate they have.
This result is closely related to the bond percolation problem.
The epidemic threshold in a susceptible-infected-recovered (SIR) model
coincides with the threshold for the bond percolation problem on
networks \cite{Newman02a}.
On the other hand,  
in a closed system of the conventional
susceptible-hidden-infected-recovered
(SHIR) or SIR model,
the number of infection
is initially increased and saturated, finally converged to zero as the
extinction. 
The pattern may be different in an open system, in fact,  
oscillations have been described 
by a deterministic Kermack-McKendrik model \cite{Shigesada97}. 
However a constant population (equal rates of the birth and the death)
or territorial competition has been mainly discussed in the model,
the growth of computer network or world-wide human contacts  
is obviously more rapid,  
and the communications of mailing or the transportations  
are not competitive. 
Even in a new report for modeling the SARS epidemic \cite{Lipsitch03}, 
the mechanism of widely rapid spreading is not clear enough,  
though further treatments for the heterogeneity in transmission through  
superspreaders has been pointed out 
and the effectiveness of quarantine has been discussed. 
The strategic quarantine in both the model and experiences in Hong Kon  
is potentially related to prevent the  
typically observed  
recoverable prevalence \cite{Kephart93b} \cite{Trendmicro03} 
in an open system as growing with new individuals.  
Fig. \ref{fig_NIMDA} shows an example of typically observed pattern in the
prevalence of computer viruses.

Thus, we consider a growing system on heterogeneous SF networks for
e-mails or human contacts,
and investigate the mechanisms of recoverable prevalence and extinction 
of infection.  
In the previous paper \cite{Hayashi03}, 
we have derived the conditions of extinction for the spreading in  
deterministic SIR 
models on SF networks by a mean-field approximation without the 
connectivity correlations. 
Although the connectivity correlations are not found  
in all growing network models and real systems \cite{Satorras01b}, 
they are at least quantitatively  
significant for the spreading on heterogeneous SF networks 
\cite{Boguna02} \cite{Callaway01} \cite{Krapivsky01}. 
In this paper, we extend the previous results to correlated cases 
between the degree of connections, and discuss the mechanisms of 
recoverable prevalence and extinction on linearly growing SF networks.

Heterogeneous SIR model.$-$On heterogeneous SF networks,  
we consider a deterministic SIR model \cite{Moreno02} 
with macroscopic equations for the number of states.  
The state transition is from susceptible, infected, to  
recovered or immune, whose numbers for each connectivity $k$ are denoted 
by  
$S_{k}(t) > 0$, $I_{k}(t) > 0$, and $R_{k}(t) > 0$, respectively. 
Microscopically, susceptible nodes stochastically 
become the infected by contacts with 
infected nodes, which are recovered by the detection or immunization. 
Assuming that infection sources exist in an initial small network  
and that both network growth and the spread of viruses  
are simultaneously progressed, we introduce a linear kernel  
\cite{Krapivsky01} as $N_{k}(t) \sim a_{k} \times t$, 
$N_{k}(t) = S_{k}(t) + I_{k}(t) + R_{k}(t)$, the growth rate  
$a_{k} \stackrel{\rm def}{=} A k^{-\nu}$, $A > 0$, $\nu > 2$. 
Note that the total  
$N(t) = \sum_{k} N_{k}(t) \sim \left( \sum_{k} a_{k} \right) \times t$ 
means a linear growth of network size. 
Since the maximum degree increases as progressing the time 
and approaches to infinity, 
it has a nearly constant growth rate  
$\sum_{k = m}^{\infty} a_{k} \sim \int_{m}^{\infty} A k^{- \nu} d k 
= \frac{A m^{1 - \nu}}{\nu - 1}$
for large $t$.
As shown in \cite{Krapivsky01}, the introduction of linear kernel 
is not contradiction with the 
preferential (linear) attachment \cite{Albert01} \cite{Barabasi99}.

We first consider a simple case with only the detection of viruses.
The time evolutions of $S_{k} > 0$ and $I_{k} > 0$ 
for each class of the connectivity $k$ are given by 
\begin{eqnarray}
  \frac{d S_{k}(t)}{dt} & = & 
	-b k S_{k}(t) \Theta_{k}(t) + a_{k}, 
	\label{eq_hete_S_recover} \\
  \frac{d I_{k}(t)}{dt} & = & 
	-\delta_{0} I_{k}(t) + b k S_{k}(t) \Theta_{k}(t),
	\label{eq_hete_I_recover}
\end{eqnarray}
where $b$ and $\delta_{0}$ 
denote the infection and detection rates between $0$ and $1$, 
the shadow variable $R_{k}(t)$ is implicitly defined  
by $\frac{d R_{k}(t)}{d t} = \delta_{0} I_{k}(t)$.
The factor  
$\Theta_{k}(t) \stackrel{\rm def}{=} \sum_{l} \frac{n_{kl}}{n_{k}} I_{l}(t)$
represents the expectation of infection by contacts from degree classes  
$\{ l \}$ to the degree class $k$.
We assume the correlation between degrees $n_{k l} \neq n_{l k}$,  
$n_{k} \stackrel{\rm def}{=} \sum_{l} n_{k l}$,
and $N_{k l} \sim n_{k l} \times t$ which is defined by the number of
nodes with degree $k$ attached to a node with degree $l$
at the time $t$.
We can easily check $\frac{d N_{k}}{d t} = a_{k}$ and the solution 
$N_{k}(t) = N_{k}(0) + a_{k} \times t \sim a_{k} \times t$  
because of the initial small size (almost all $N_{k}(0) = 0$ except 
for a few small $k$). 
The network is keeping the power-law degree distribution  
$P(k) \sim k^{- \nu}$ in growing. 
Although the growing in our model is not exactly same in 
Krapivsky-Render's GN model \cite{Krapivsky01},
the growth rate $a_{k}$ is close to their 
$n_{k} = \frac{4}{k (k+1) (k + 2)} \sim 4 k^{-3}$ 
in large $t$.

\begin{figure}[htb]
  \begin{minipage}{.47\textwidth}
    \includegraphics[width=47mm]{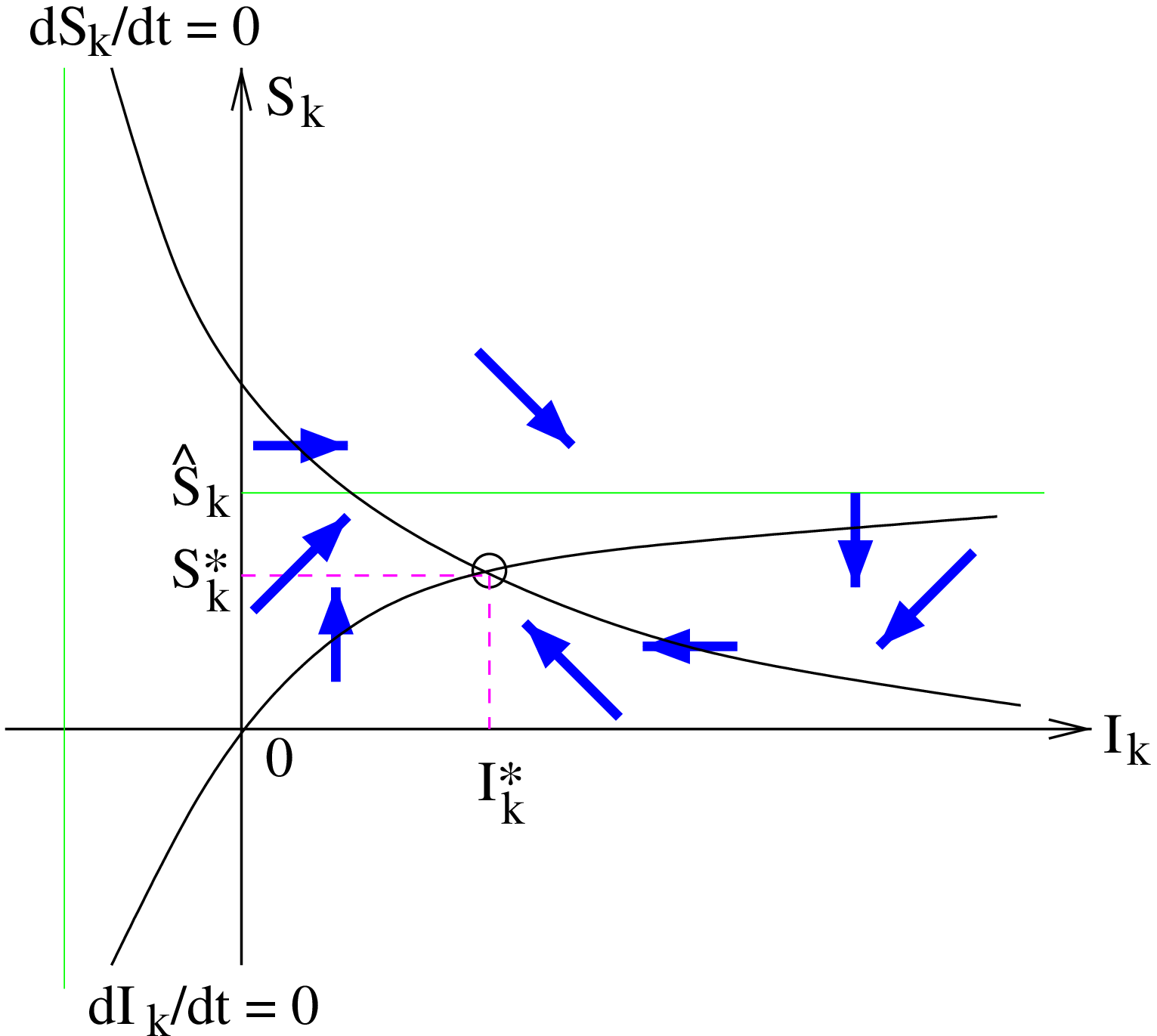} 
    \begin{center} (a) \end{center}
  \end{minipage} 
  \hfill 
  \begin{minipage}{.47\textwidth}
    \includegraphics[width=47mm]{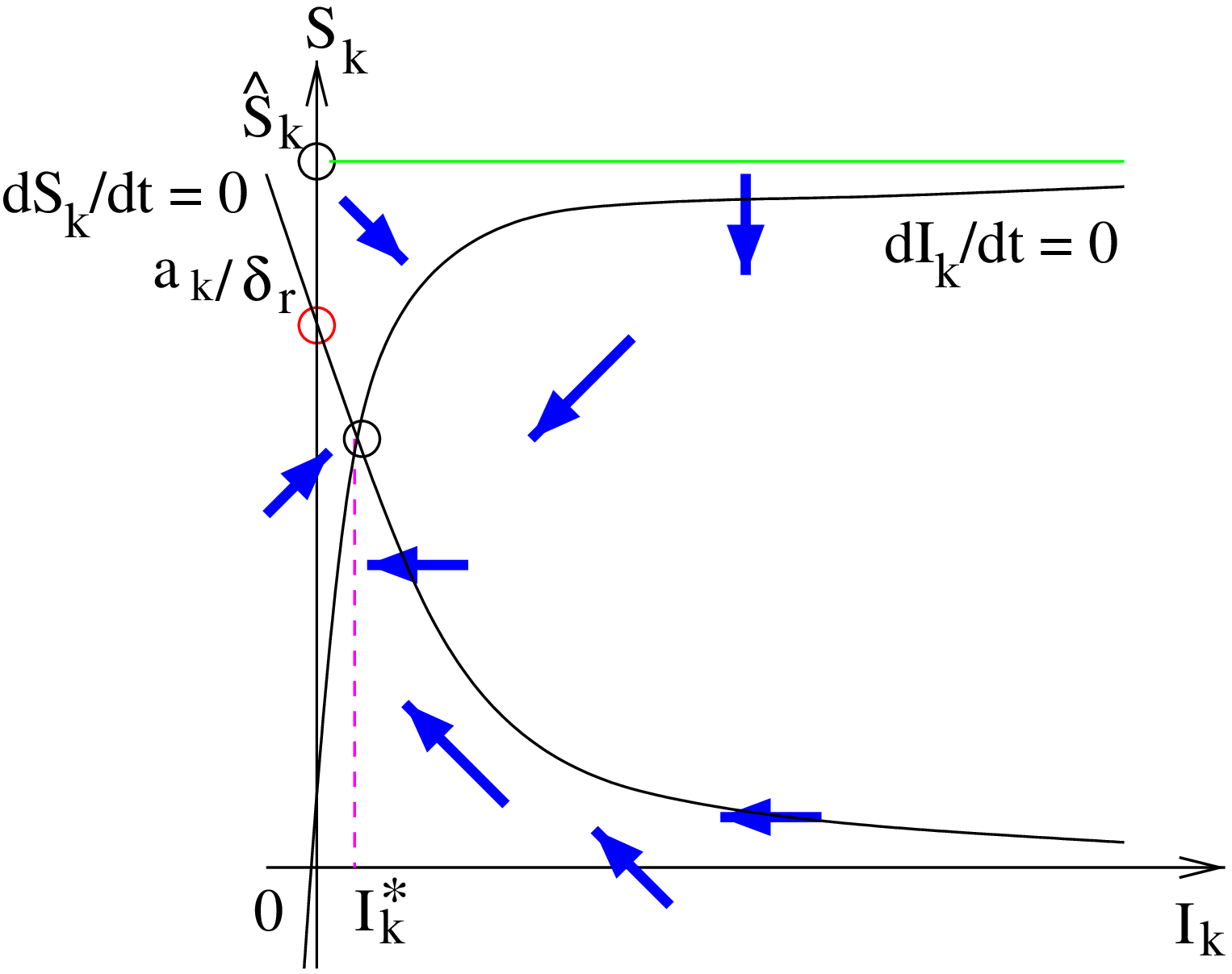} 
    \begin{center} (b) \end{center}
  \end{minipage} 
  \hfill 
  \begin{minipage}{.47\textwidth}
    \includegraphics[width=47mm]{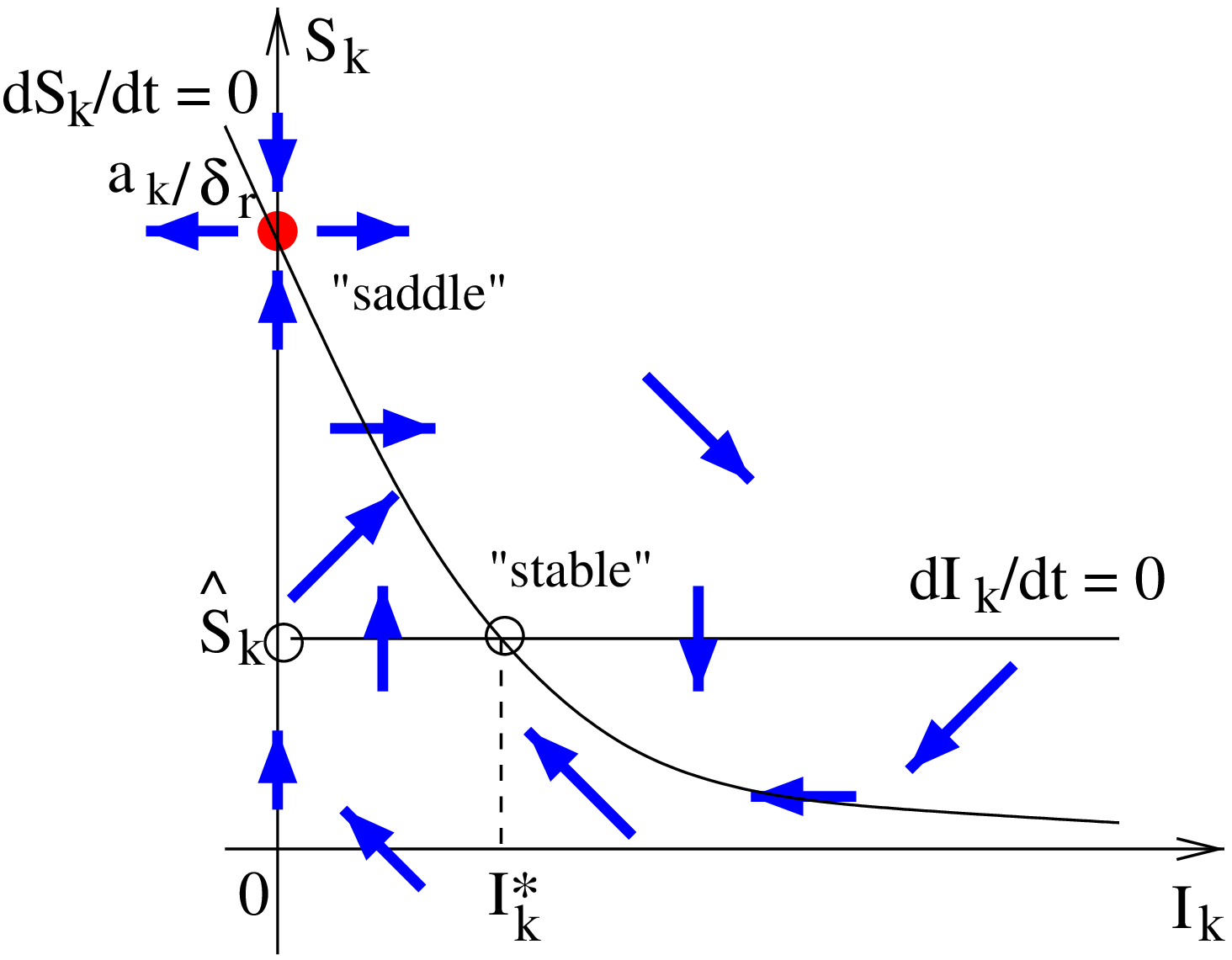} 
    \begin{center} (c) \end{center}
  \end{minipage} 
  \hfill 
  \begin{minipage}{.47\textwidth}
    \includegraphics[width=47mm]{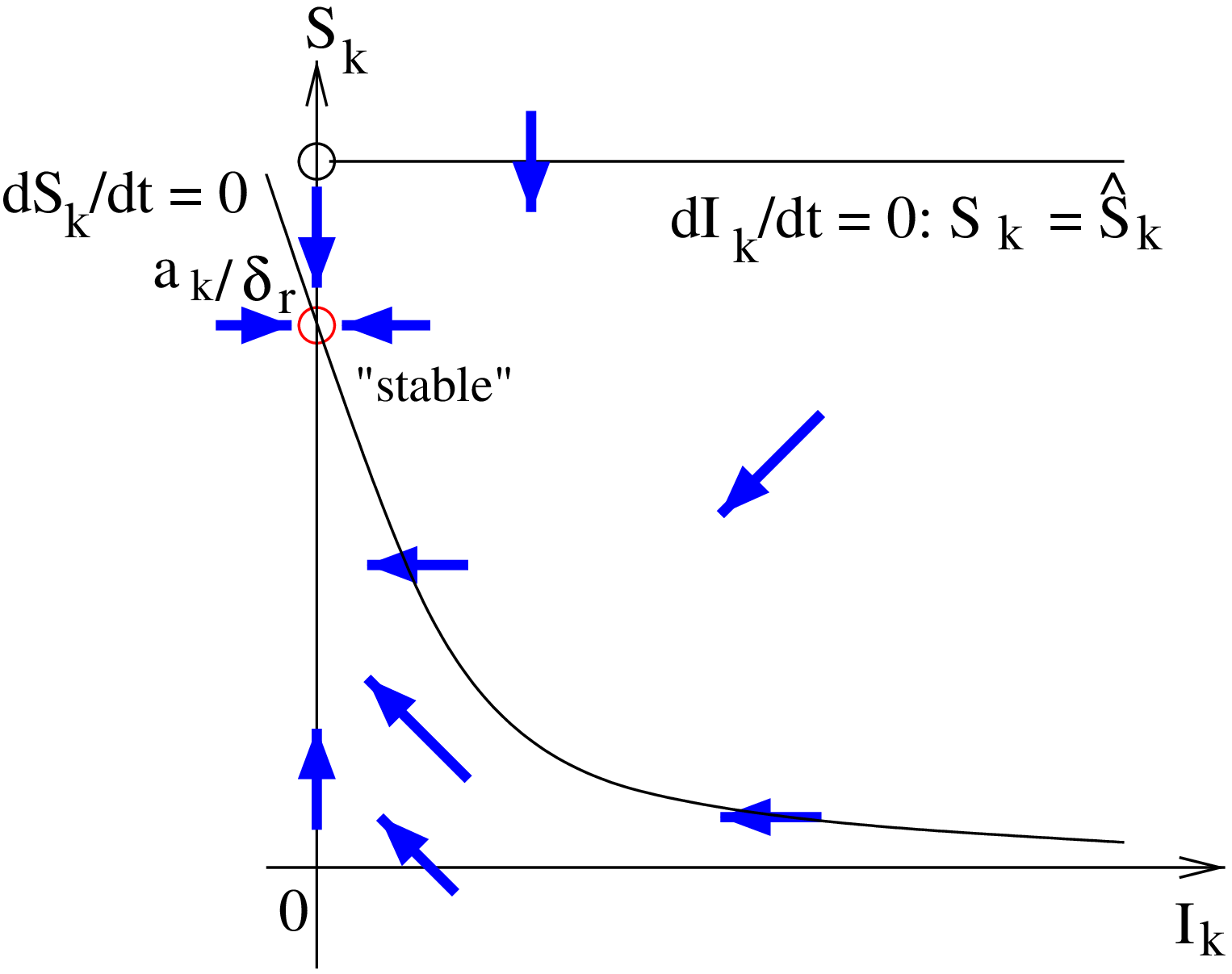} 
    \begin{center} (d) \end{center}
  \end{minipage} 
  \caption{Nullclines and the vector fields near an equilibrium point. 
(a) damping oscillation on a section $I_{l}^{*}$: const. for all $l \neq k$, 
(b) nearly extinction on a section $I_{l}^{*} \approx 0$, (c) divergence from a saddle, and (d) convergence to the extinction 
$I_{k}^{*} \rightarrow 0$ on a section $I_{l}^{*} = 0$.}
  \label{fig_vector_SIR}
\end{figure}

We consider a section of $I_{l} = I^{*}_{l}$: const. 
for all $l \neq k$. 
Fig. \ref{fig_vector_SIR} (a) shows the nullclines of
\[
  \frac{d S_{k}}{dt} = 0: \;
	S_{k} = \frac{a_{k}}{k b \Theta_{k}}
	= \frac{a_{k} n_{k}}{k b (n_{kk} I_{k} 
	+ \sum_{l \neq k} n_{kl} I_{l}^{*})},
\]
\[
  \frac{d I_{k}}{dt} = 0: \;
	S_{k} = \frac{\delta_{0} I_{k}}{k b \Theta_{k}}
	= \frac{n_{k} \delta_{0} I_{k}}{k b (n_{kk} I_{k} 
	+ \sum_{l \neq k} n_{kl} I_{l}^{*})}, 
\]
and the vector filed for Eqs.
(\ref{eq_hete_S_recover})(\ref{eq_hete_I_recover}). 
There exist an equilibrium point:  
$I_{k}^{*} = \frac{a_{k}}{\delta_{0}} = \frac{A}{\delta_{0} k^{\nu}}$, 
$S_{k}^{*} = \frac{a_{k}}{b k \Theta_{k}^{*}} \sim \frac{A^{2} 
\delta_{0}}{b k^{2 \nu +1} \sum_{l} n_{kl} l^{-\nu}}$.
On the state space, the viruses are not extinct 
by only the detection in even $I_{l}^{*} = 0$ ($l \neq k$),  
unless the growing is stopped as $a_{k} = 0$.

Effect of immunization.$-$Next, we 
study the effect of random and hub immunization. 
With the randomly immune rate $0 < \delta_{r} <1$,
the time evolutions are given by 
\begin{eqnarray}
  \frac{d S_{k}(t)}{dt} & = & 
	-b k S_{k}(t) \Theta_{k}(t) + a_{k} - \delta_{r} S_{k}(t),
	\label{eq_hete_S_immune}\\
  \frac{d I_{k}(t)}{dt} & = & 
	-\delta_{0} I_{k}(t) + b k S_{k}(t) \Theta_{k}(t)
	- \delta_{r} I_{k}(t),
	\label{eq_hete_I_immune}
\end{eqnarray}
where the shadow variable $R_{k}(t)$ is also defined by 
$\frac{d R_{k}(t)}{d t} 
	= \delta_{0} I_{k}(t) + \delta_{r} (S_{k}(t) + I_{k}(t))$.

On the section by assuming $\exists I_{l}^{*} > 0$ for $l \neq k$, 
the nullclines are 
\[
  \frac{d S_{k}}{dt} = 0: \;
	S_{k} = \frac{a_{k}}{\delta_{r} + k b \Theta_{k}},
\]
\[
  \frac{d I_{k}}{dt} = 0: \;
	S_{k} = \frac{(\delta_{0} + \delta_{r}) I_{k}}{b k 
	\Theta_{k}}, 
\]
for Eqs. (\ref{eq_hete_S_immune})(\ref{eq_hete_I_immune}).
At the intersection of nullclines, from 
$I_{k}^{*}
	= \frac{a_{k} - \delta_{r} S_{k}^{*}}{\delta_{0} + \delta{r}} 
	= \frac{a_{k}}{\delta_{0} + \delta{r}} \left( 
	1 - \frac{\delta_{r}}{\delta_{r} + k b \Theta_{k}^{*}} \right)$,
the self-consistent solution is given by 
\begin{equation}
 \begin{array}{ll}
 \Theta_{k}^{*} & = \sum_{l} \frac{n_{kl}}{n_{k}} I_{l}^{*}
 = \frac{1}{n_{k} (\delta_{r} + \delta_{0})} \\
 & \times \left\{
 \sum_{l \neq k} a_{l} n_{kl} 
 \left( 1 - \frac{\delta_{r}}{\delta_{r} + l b \Theta_{l}^{*}} \right) 
 + a_{k} n_{kk} 
 \left( 1 - \frac{\delta_{r}}{\delta_{r} + k b \Theta_{k}^{*}} \right) 
 \right\}. \label{eq_self-consistent}
 \end{array}
\end{equation}
When the right hand side of (\ref{eq_self-consistent}) is denoted by 
$f(\Theta_{k})$, for $\exists \Theta_{k} > 0$, 
$\frac{d f(\Theta_{k}))}{d \Theta_{k}}|_{\Theta_{k} = 0} > 1$ 
is necessary.
The condition is 
\begin{equation}
 \frac{d f(\Theta_{k})}{d \Theta_{k}}\Big{|}_{\Theta_{k} = 0} = 
  \frac{k a_{k} b \times n_{kk}}{n_{k} \delta_{r} 
  (\delta_{r} + \delta_{0})} > 1. \label{eq_cond_f}
\end{equation}
This means the growth and infection rates should be larger than the
immune and detection rates according to the connectivity 
correlation of degree $k$.

On the other hand, 
we assume $I^{*}_{l} = 0$ for $l \neq k$ 
to discuss the extinction. 
The necessary condition of extinction is given by that 
the point $(0, \frac{a_{k}}{\delta_{r}})$ on the hyperbolic nullcline of 
$\frac{d S_{k}}{dt} = 0$ is below the line 
$S_{k} = \hat{S}_{k} \stackrel{\rm def}{=} 
\frac{(\delta_{0} + \delta_{r}) n_{k}}{k b n_{kk}}$: const. 
of $\frac{d I_{k}}{dt} = 0$. 
From the condition 
\begin{equation}
 \frac{a_{k}}{\delta_{r}} < \frac{(\delta_{0} + \delta_{r}) 
n_{k}}{k b n_{kk}}, \label{eq_cond_eigen}
\end{equation}
we obtain 
\begin{equation}
 \begin{array}{lll}
  \delta_{r} & > & - \delta_{0} 
   + \sqrt{\delta_{0}^{2} + 4 k a_{k} b \times n_{kk} / n_{k}}\\
  \label{eq_cond_extinction}
  & \sim & - \delta_{0} + \sqrt{\delta_{0}^{2} + 4 k b \times n_{kk}},
 \end{array}
\end{equation}
for $n_{k} \sim a_{k}$ asymptotically.
From $\delta_{r}< 1$, we further derive 
\[
 A < \min_{k \geq m} \left\{ \frac{ (1 + 2 \delta_{0}) n_{k} 
m^{\nu -1}}{b n_{kk}} \right\},
\]
as a limitation of growth rate or a weak correlation
\[
 n_{kk} < \frac{1 + 2 \delta_{0}}{4 k b},
\]
for the extinction.
In this extinction case, 
there exists a stable equilibrium point, 
otherwise a saddle and a stable equilibrium point
as shown in Figs. \ref{fig_vector_SIR} (c)(d).
Note that $\delta_{r} > 0$ is important for the extinction.
When no growing $a_{k} = 0$ (as a closed system) is applied to the above
results, the condition (\ref{eq_cond_f}) for damping
oscillation is not satisfied, 
while the extinction condition (\ref{eq_cond_extinction}) is satisfied.
In other words, it is suggested that 
a growing network causes the recoverable prevalence of infection.

By replacing $a_{k} = A k^{-\nu}$ with the Krapivsky-Render's $n_{k}$
\cite{Krapivsky01} and using their 
$n_{kk} = \frac{(k - 1)(5 k + 2)}{k^{2} (k + 1)^{2} (2 k + 1) (2 k -1)}$, 
$k \geq 2$, we can check that 
$n_{kk} < \frac{1 + 2 \delta_{0}}{4 k b}$ is satisfied 
for any $0 < b, \delta_{0} < 1$.
The reason why viruses tend to be extinct
in their GN model by even random immunization 
is that the infected parts are disconnected by
recovered nodes on the tree structure,
while our model generally has other bypass links than the links of tree in the
difference of the degree distributions $a_{k} t$ 
and their $n_{k} t$ for small $k$, therefore 
infection can be spreading.

Moreover, for the hub immunization \cite{Dezso02}, 
$\delta_{r}$ is replaced by $0 < \delta_{h} k^{\tau} < k^{\tau}$, 
$\tau > 0$.
Then, the necessary condition of extinction in (\ref{eq_cond_extinction})
is relaxed to larger 
\[
 A < \min_{k \geq m} \left\{ \frac{ (m^{\tau} + 2 \delta_{0}) n_{k} 
m^{(\nu + \tau -1)}}{b n_{kk}} \right\},
\]
\[
 n_{kk} < \frac{m^{\tau}(m^{\tau} + 2 \delta_{0})}{4 k b}.
\]
However this is under the assumption of $I_{l}^{*} = 0$.
If the general case for a section $I^{*}_{l} \neq 0$ is
similar to Fig. \ref{fig_vector_SIR} (a), 
then viruses in the class of degree $k$
are not extinct unless the growing is stopped as $a_{k} = 0$,
otherwise they approaches to the extinction as shown from
Fig. \ref{fig_vector_SIR} (b) to (d).
In fact, at the point 
$(\ldots, I_{k}, S_{k}, \ldots, I_{l}, S_{l}, \ldots) = (\ldots, 0,
\frac{a_{k}}{\delta_{r}}, \ldots, 0, \frac{a_{l}}{\delta_{r}}, \ldots)$, 
the eigenvalues of Jacobian 
\[
 \left( 
  \begin{array}{ccccccc}
   \ddots & \ldots & \ldots & \ldots & \ldots & \ldots  & \ldots\\
   \ldots & -\delta_{r} & -c_{kk} & \ldots & 0 & -c_{kl} & \ldots\\
   \ldots & 0 & d_{kk} 
    & \ldots & 0 & c_{kl} & \ldots\\
   \ldots & \ldots & \ldots & \ddots & \ldots & \ldots  & \ldots\\
   \ldots & 0 & -c_{lk} & \ldots & -\delta_{r} & -c_{ll} & \ldots\\
   \ldots & 0 &  c_{lk} & \ldots & 0 & d_{ll} 
    & \ldots\\
   \ldots & \ldots & \ldots & \ldots & \ldots & \ldots  & \ddots
  \end{array} 
 \right)
\]
are $\lambda = - \delta_{r} < 0$ and 
$\lambda \sim d_{kk} < 0$
from the condition (\ref{eq_cond_eigen}),
where 
$d_{kk} \stackrel{\rm def}{=} -(\delta_{0} + \delta_{r}) + 
\frac{k a_{k} b n_{kk}}{\delta_{r} n_{k}}$
and 
$c_{kl} \stackrel{\rm def}{=} 
\frac{k a_{k} b n_{kl}}{\delta_{r} n_{k}} \sim \frac{k b n_{kl}}{\delta_{r}}$,
if $c_{kl}$ and $c_{lk}$ with respect to the cross correlations 
are negligibly weak.
Another extinction route is given by $b \rightarrow 0$, in which case 
the condition (\ref{eq_cond_extinction})
is satisfied for any $A$ and $n_{kk}$.
It is consistent with the result of
no epidemic threshold \cite{Satorras01a}.

These dynamics are qualitatively same to the previous results by a
mean-field approximation without the connectivity correlations
\cite{Hayashi03}.
In summary, we have investigate mechanisms of recoverable prevalence and
extinction of viruses in SIR models with the connectivity correlations 
on linearly growing SF networks,
and derived the necessary condition of extinction 
related to the growth, infection, detection, and immune rates.
The results suggest that we must stop the growing of network or 
close it by quarantine to prevent the spreading of infection.
A more general case in linearly growing SF networks on no 
assumption of correlation 
$N_{k l} \sim n_{k l} \times t$ is further study.

\end{document}